# Laundering Money Online: a review of cybercriminals' methods

Jean-Loup Richet

## Executive Summary

Money laundering is a critical step in the cyber crime process which is experiencing some changes as hackers and their criminal colleagues continually alter and optimize payment mechanisms. Conducting quantitative research on underground laundering activity poses an inherent challenge: Bad guys and their banks don't share information on criminal pursuits. However, by analyzing forums, we have identified two growth areas in money laundering:

- **Online gaming**—Online role playing games provide an easy way for criminals to launder money. This frequently involves the opening of numerous different accounts on various online games to move money.
- **Micro laundering**—Cyber criminals are increasingly looking at micro laundering via sites like PayPal or, interestingly, using job advertising sites, to avoid detection. Moreover, as online and mobile micro-payment are interconnected with traditional payment services, funds can now be moved to or from a variety of payment methods, increasing the difficulty to apprehend money launderers. Micro laundering makes it possible to launder a large amount of money in small amounts through thousands of electronic transactions. One growing scenario: using virtual credit cards as an alternative to prepaid mobile cards; they could be funded with a scammed bank account – with instant transaction – and used as a foundation of a PayPal account that would be laundered through a micro-laundering scheme.

# Laundering Money Online: a review of cybercriminals' methods

Millions of transactions take place over the internet each day, and criminal organizations are taking advantage of this fact to launder illegally acquired funds through covert, anonymous online transactions. The more robust and complex the various online marketplaces become the more untraceable methods criminals are finding to pass 'dirty' money into online accounts and pull 'clean' money out of others. The anonymous nature of the internet and the ever evolving technologies available allow numerous opportunities for online money laundering operations to take place. Many of these methods involve using a ruse to pull unsuspecting participants into their money laundering schemes, often with serious financial and legal consequences for victims. The best way for law abiding citizens to avoid becoming complicit in such illegal activities is to stay informed as to the methods criminals are using to pull them in.

We all know the oldest 'physical' placement methods of money launderers: cash smuggling, casinos and other gambling venues, insurance policies (launderers purchase them and then redeem them at a discount, paying fees and penalties but receiving a clean check from the insurance company), *hawalas / fe chi'en* or the black market peso exchange (informal value transfer system), shell corporations, and so on and so forth. But there is also a number of online money laundering schemes currently being used by criminal enterprises to pass illegally received funds through legitimate accounts, and new ones are popping up all the time. Some of the most widespread schemes are detailed in this article.

## Methodology

Ostensibly, conducting quantitative research on underground laundering activity poses an inherent challenge: Bad guys and their banks don't share information on criminal pursuits. Our approach utilizes an online ethnography, observing large online hacker forums and communities and researching topics related to money laundering on their databases. We used a large variety of keywords, from those linked with payment solutions to those associated with black markets. After a first review, we filtered our data, and discarded irrelevant forum threads. We then analyzed the content of these threads and synthesize our findings into the following categories.

# The classics: from money mule scams to black markets

## Liberty Reserve

Frequently associated with carding communities[1], Liberty Reserve (LR) had a bad reputation amongst some Black Hats boards[2].

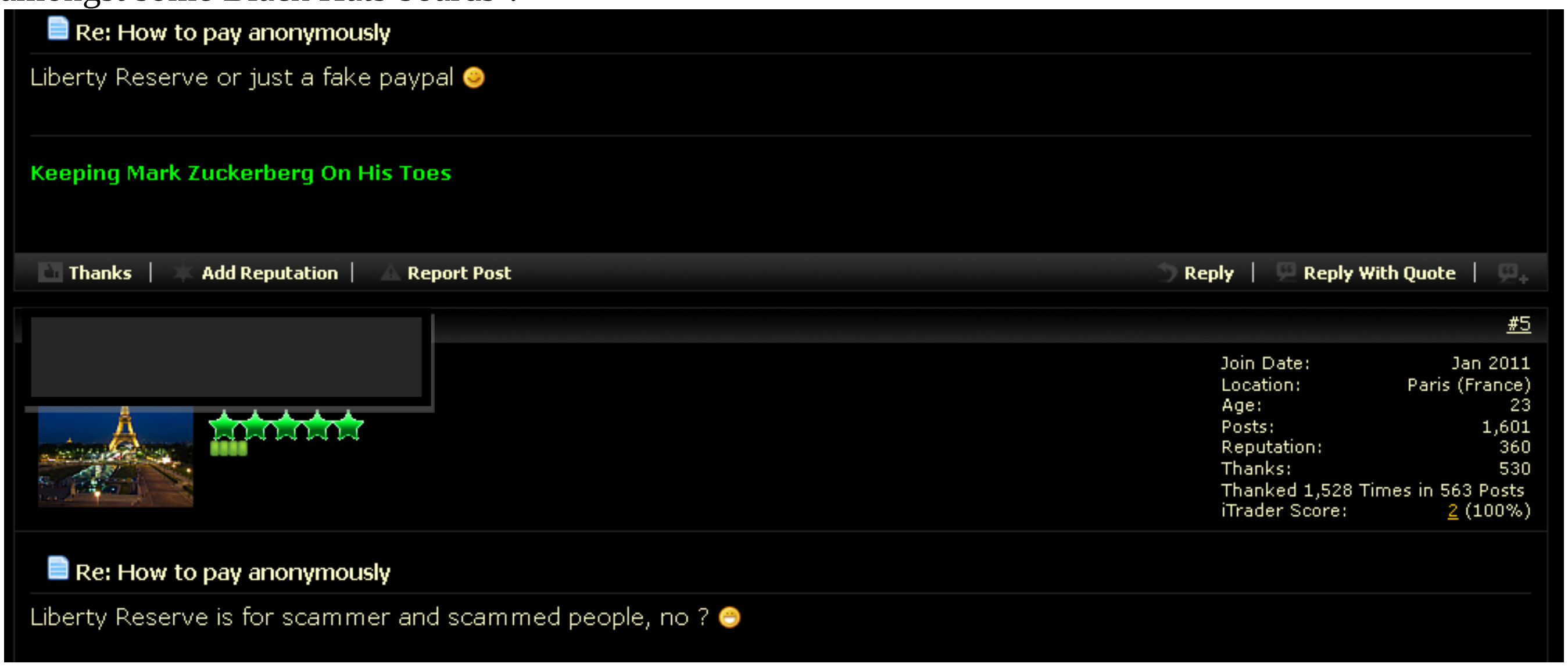

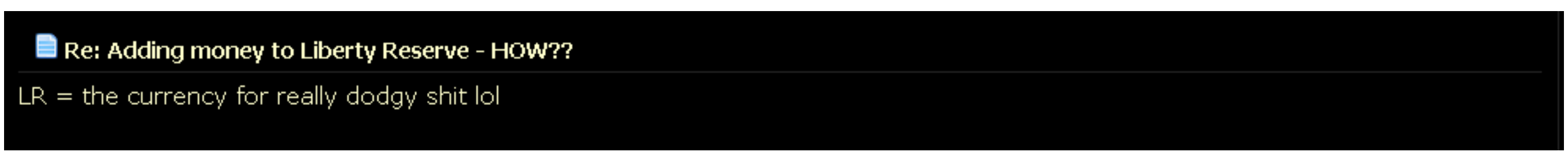

But there were very few ways to receive money anonymously, and LR was one of them. Thus LR was widely advised when one wanted to launder money:

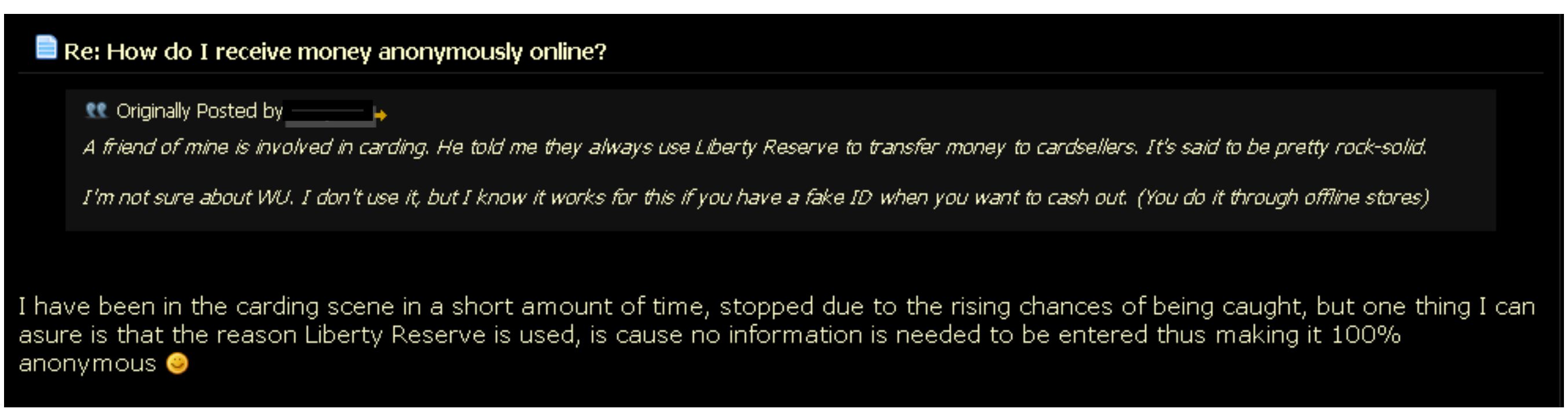

[1] Cybercriminals involved in credit card fraud.
[2] Internet forum boards which aim to share advice on cybercrime and fraudulent methods.

> **Re: The Best Payment Processors?**
>
> The best patment processors is liberty reserve .......its too safe....

> **Re: Best method for transferring money that's not paypal?**
>
> I would suggest using liberty reserve...
>
> Once the money is sent to you, it's sent to you, period. And if you're selling some blackhat stuff it's anonymous.

> Actually, no one can send money directly to / from them.. I guess thats the beauty of what makes it 100% anonymous... The dont have any of your info.. Only the exchangers you use.
>
> So when someone says: I sent $500 to account 324j5234 ... no one is going to know who that is. (and they won't know what exchanger you used of course to withdraw the money)

## How did it work?

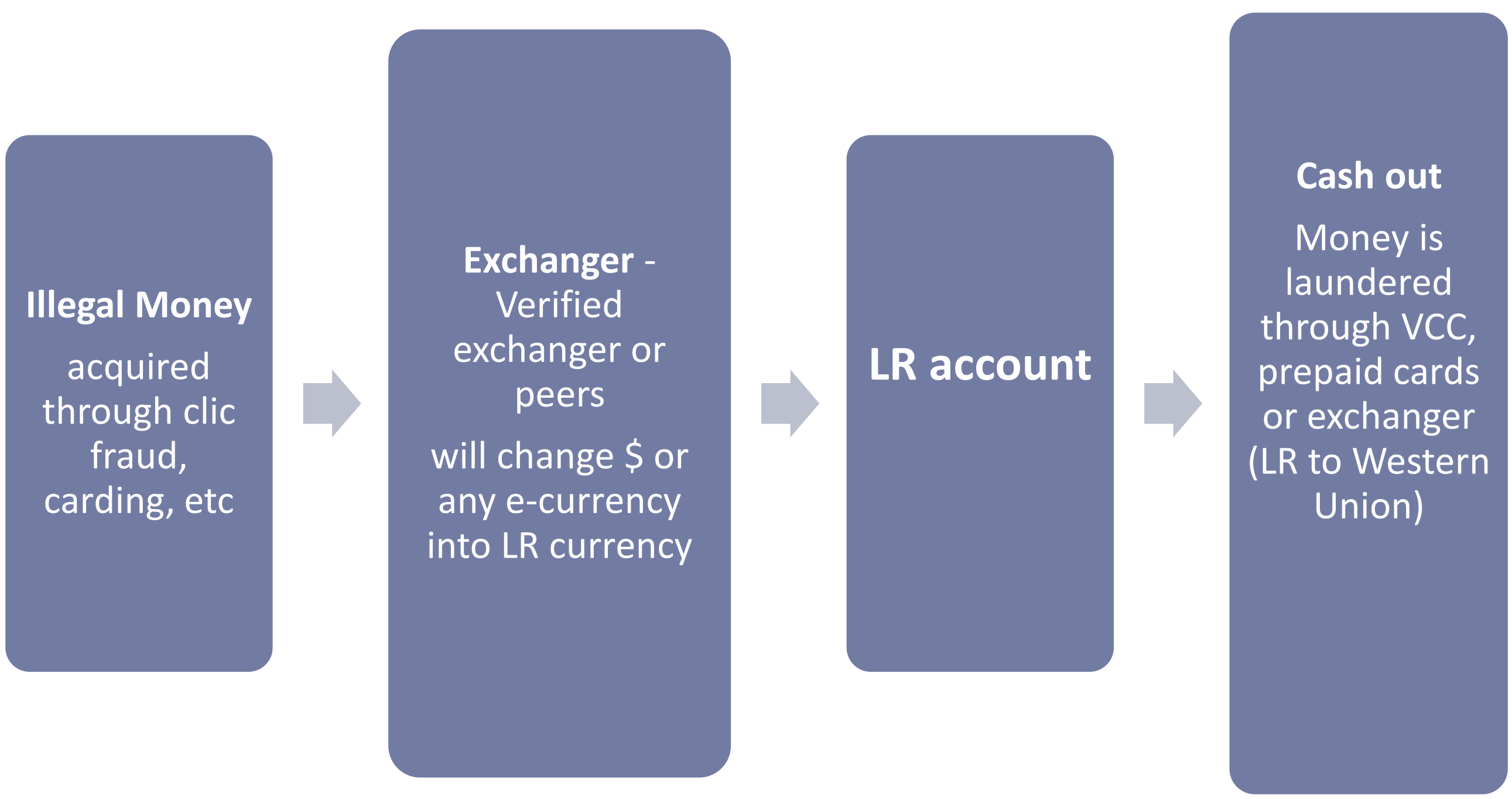

Exchanging money could be done through the use of verified exchangers such as ebuygold.com - they have been verified by Liberty Reserve. There is a lot of fly-by-night firms and scammers in this shady market. Because of that, peers markets are growing. The following screenshot is an example of a peers market (e-currency auction service, member to member), a marketplace for anyone who wants to exchange currencies..

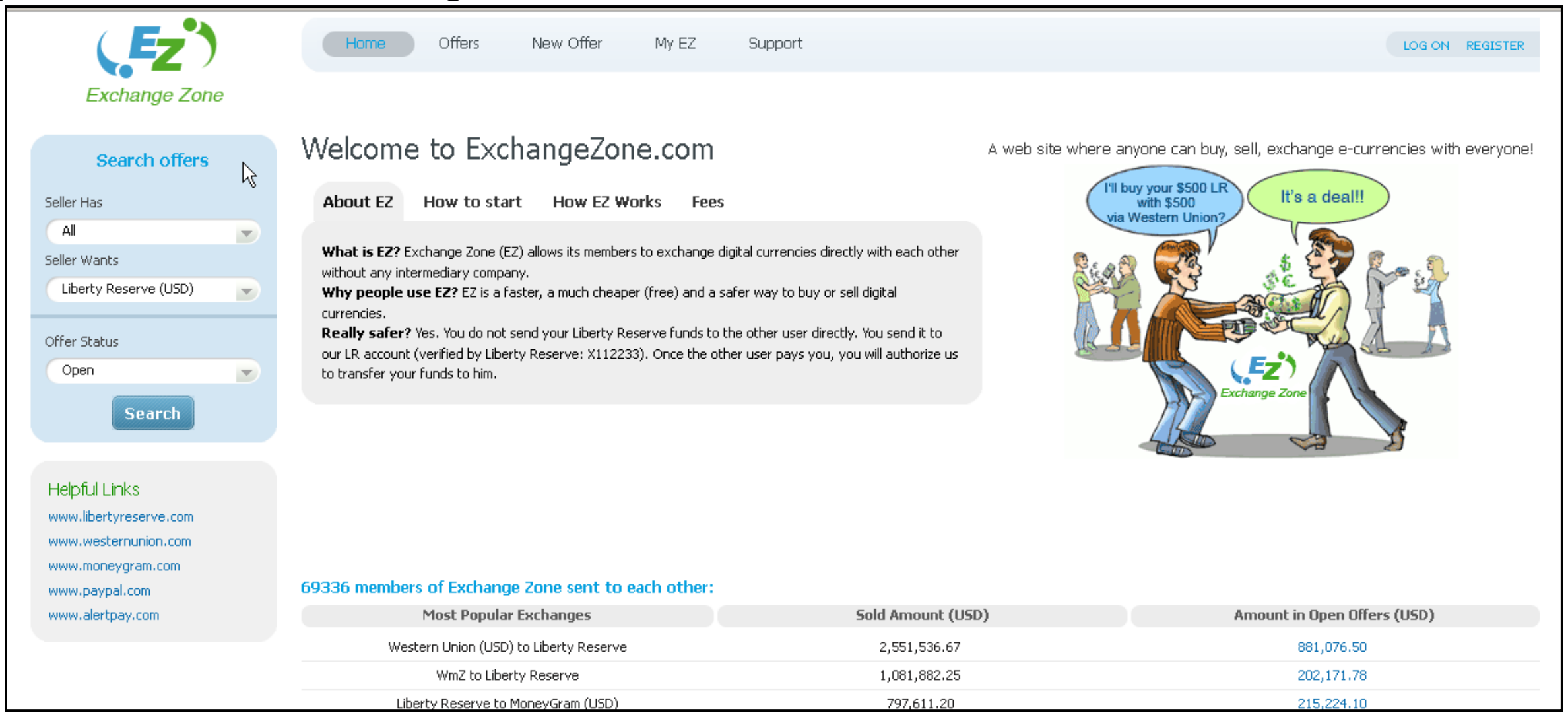

However, the closure and seizure of Liberty Reserve will not end these fraudulent practices – there are numerous competitors and alternatives (WebMoney, Bitcoins, Paymer, PerfectMoney and so on)…

## Money mule scams

**CHAPTER 1 : The Drop ( Mules)**

2 solutions is offer to u :

1: U have money to invest :

Buy some fake ID + electricity bills ... ETC ( whats ur bank ask for open an account )
Better look like a businessman ( class suits , perfume , watches ) but the attitude do all . like that u could ask good card with ur account +better limit on wildraw .

2: Without cash to invest :

Find someone who can accept a transfer on his account ( at his own risk ) or find a reliable MULE bank ( always control this guy ) .
in this case its better use someone who could control it ( ppl in real life , not e-cashier ripper drop ) *

**CHAPTER 2 : The Transfer Online ( or phone)**

In this case ; u need have login Bank ( better have same as ur drop )

Then collect a max of info for this guy . ssn , dob , mmn , adress , phone ....
+ search on all his account his life .
I explain lhe last point : see what he paid , wich bills .... ETC if he have some LOAn .....
Who have acces to this account ......

+ u have some personnal info better u could talk and act like the owner of the account hacked if u need call to the bank for change some detail ( phone ... confirm transfer .... )

*Guide posted on a board about how to use a money mule to launder hacked bank account*

This is a method you're probably familiar with if you check regularly your spam e-mail box. It involves a very friendly e-mail from someone from a foreign country claiming that they need your help to transfer a large sum of money to your country. For your help with the transfer you will receive a percentage of the transfer, which, according to the e-mail, will amount to several thousand dollars or more. All you have to do is supply your bank account information.

Some of the senders claim to be Princes, government agents, or other high ranking officials from the country they claim they are from. They are written in a very gracious and humble tone, and very politely ask for your help, as if you'd be doing them a huge favor.

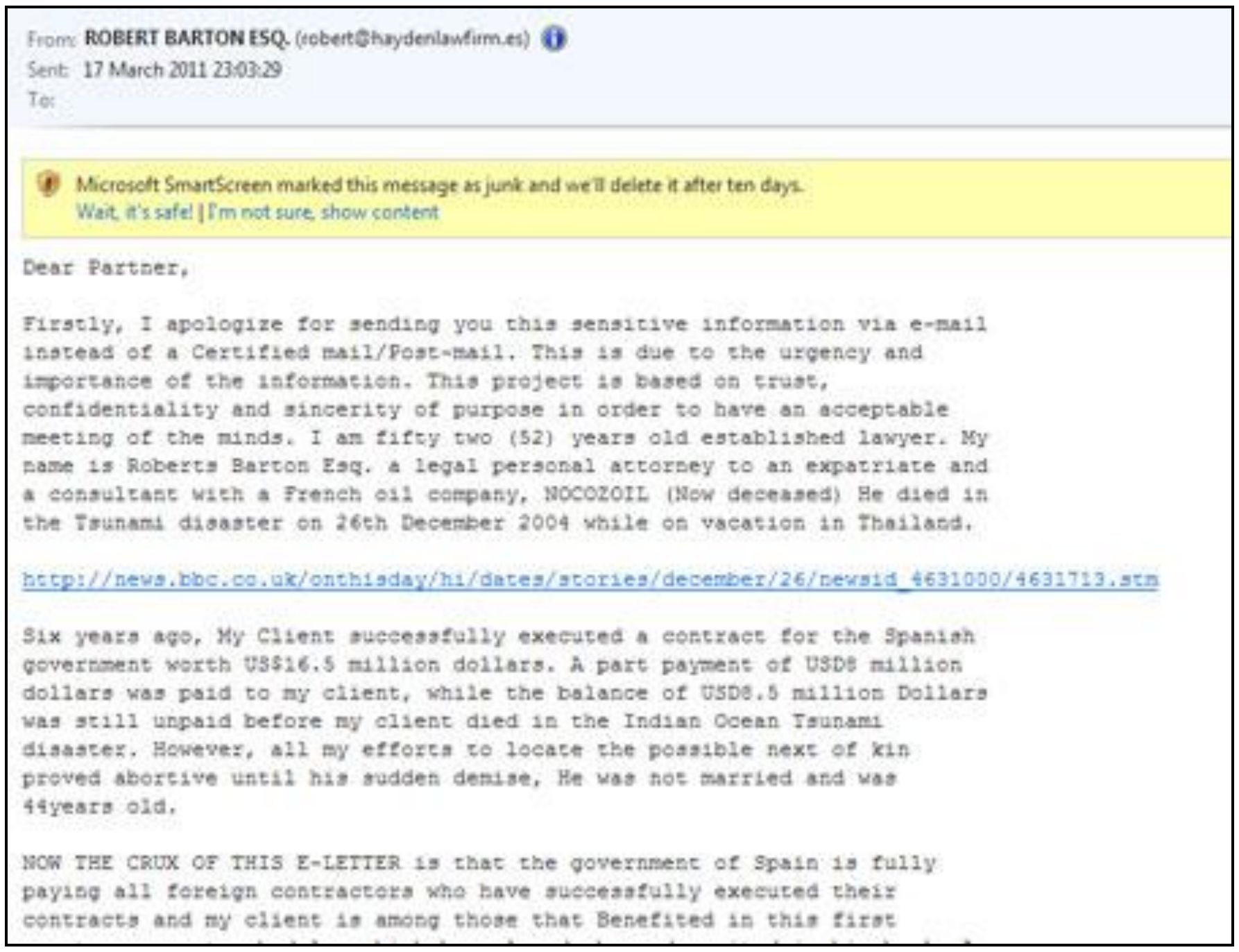
From: ROBERT BARTON ESQ. (robert@haydenlawfirm.es)
Sent: 17 March 2011 23:03:29
To:

Microsoft SmartScreen marked this message as junk and we'll delete it after ten days.
Wait, it's safe! | I'm not sure, show content

Dear Partner,

Firstly, I apologize for sending you this sensitive information via e-mail instead of a Certified mail/Post-mail. This is due to the urgency and importance of the information. This project is based on trust, confidentiality and sincerity of purpose in order to have an acceptable meeting of the minds. I am fifty two (52) years old established lawyer. My name is Roberts Barton Esq. a legal personal attorney to an expatriate and a consultant with a French oil company, NOCOZOIL (Now deceased) He died in the Tsunami disaster on 26th December 2004 while on vacation in Thailand.

http://news.bbc.co.uk/onthisday/hi/dates/stories/december/26/newsid_4631000/4631713.stm

Six years ago, My Client successfully executed a contract for the Spanish government worth US$16.5 million dollars. A part payment of USD8 million dollars was paid to my client, while the balance of USD8.5 million Dollars was still unpaid before my client died in the Indian Ocean Tsunami disaster. However, all my efforts to locate the possible next of kin proved abortive until his sudden demise, He was not married and was 44years old.

NOW THE CRUX OF THIS E-LETTER is that the government of Spain is fully paying all foreign contractors who have successfully executed their contracts and my client is among those that Benefited in this first

While most people know that any inquiry requiring your personal financial information should immediately throw up a red flag, some folks do fall victim to this kind of scam. While some of these scammers will simply try to steal money from your account, others will attempt to transfer large sums of money stolen from other accounts to you. They'll have you send the majority of the funds to an alias account of theirs in exchange for your cut. Once the bank realizes the transaction was illegal, you will be held accountable, as the stolen funds were withdrawn from your account.

Other scams involve supposed job offers. With so many individuals interested in work from home jobs, there are a countless number of potential victims here. The 'job' that you'll be doing involves accepting money transfers to your account and passing the money on to alias accounts set up by your so called employer. Of course, the transfers you'll receive will contain illegally obtained funds, and when law enforcement traces the funds, it will lead them to you, not the criminals who illegally obtained the money.

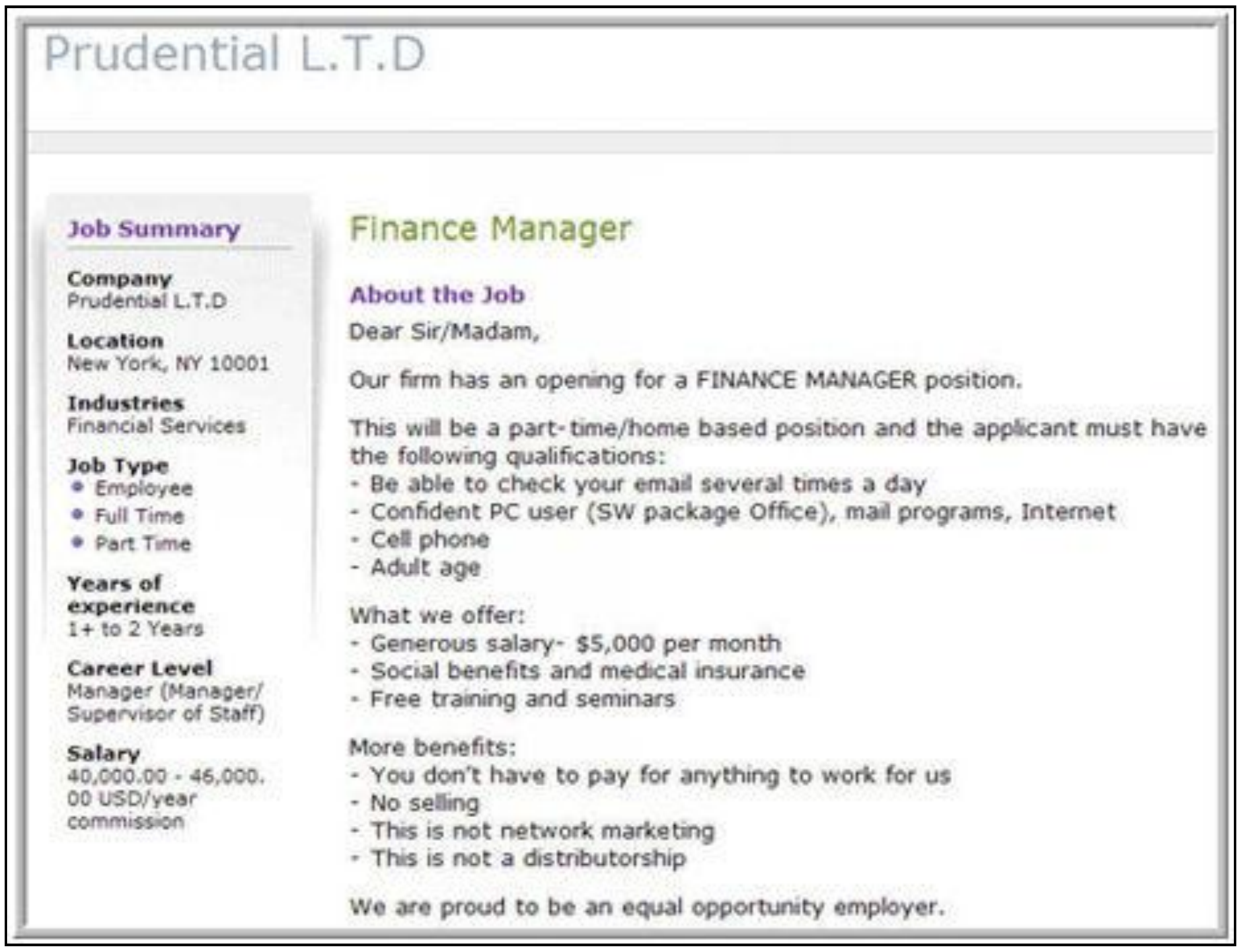

## Black Market Peso Exchange

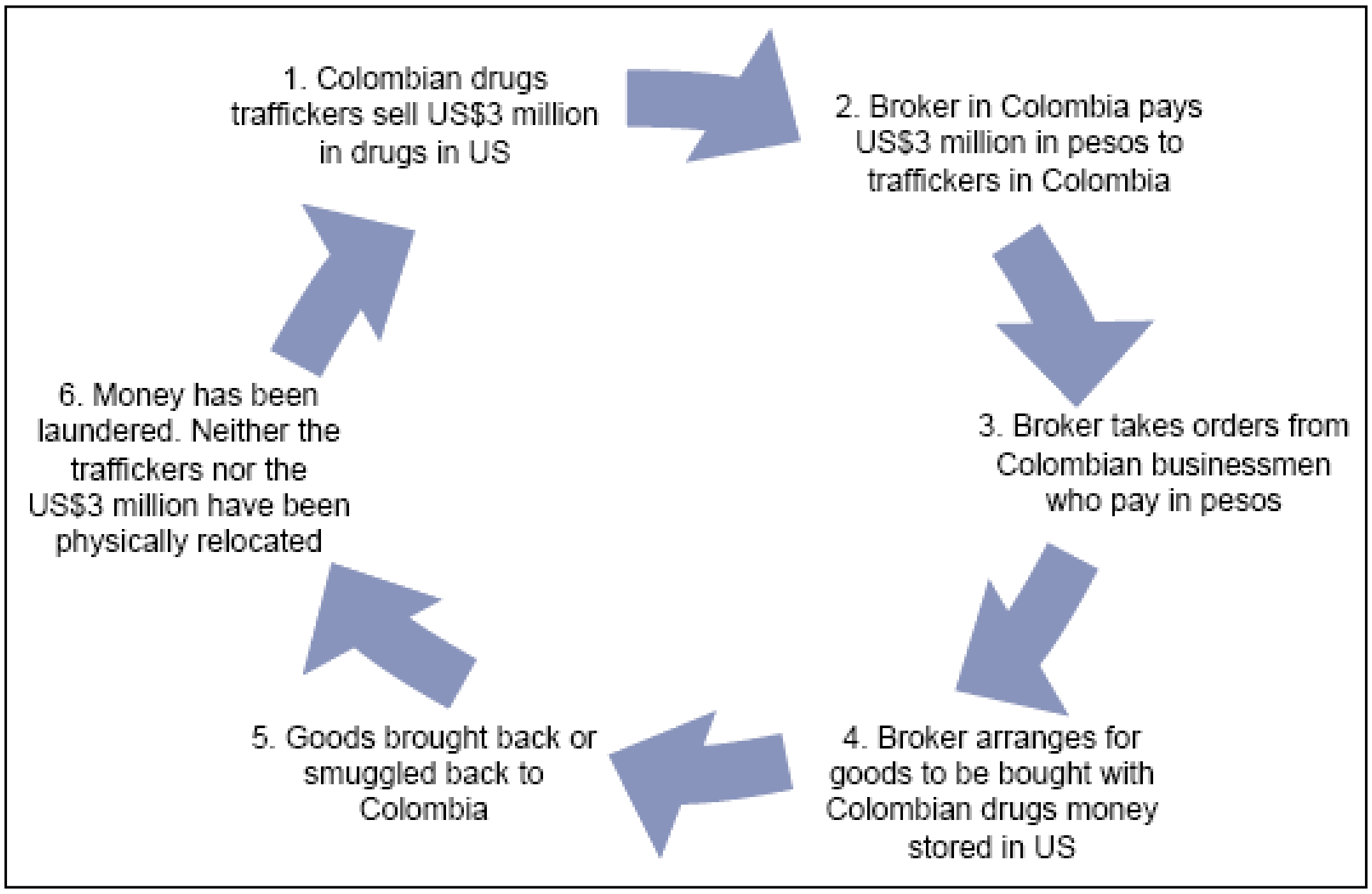

*Source: Australian Institute of Criminology*

Over $7 billion each year is laundered through Colombian corporations from Mexican and Colombian drug cartels. This is done through the Black Market Peso Exchange or BMPE. The BMPE is actually a covert system of banking that helps drug dealers to exchange American dollars for pesos. These dollars are then purchased by Colombian businessmen and used to buy American goods which are then sold back home in Colombia. This traditional form of crime is enhanced with ICT: Drugs could be sold through illegal online marketplaces such as the **Silk Road,** in e-currencies like bitcoins in order to increase the difficulty to trace the operation, while brokers use blackberry messengers and negotiations are conducted through the internet.

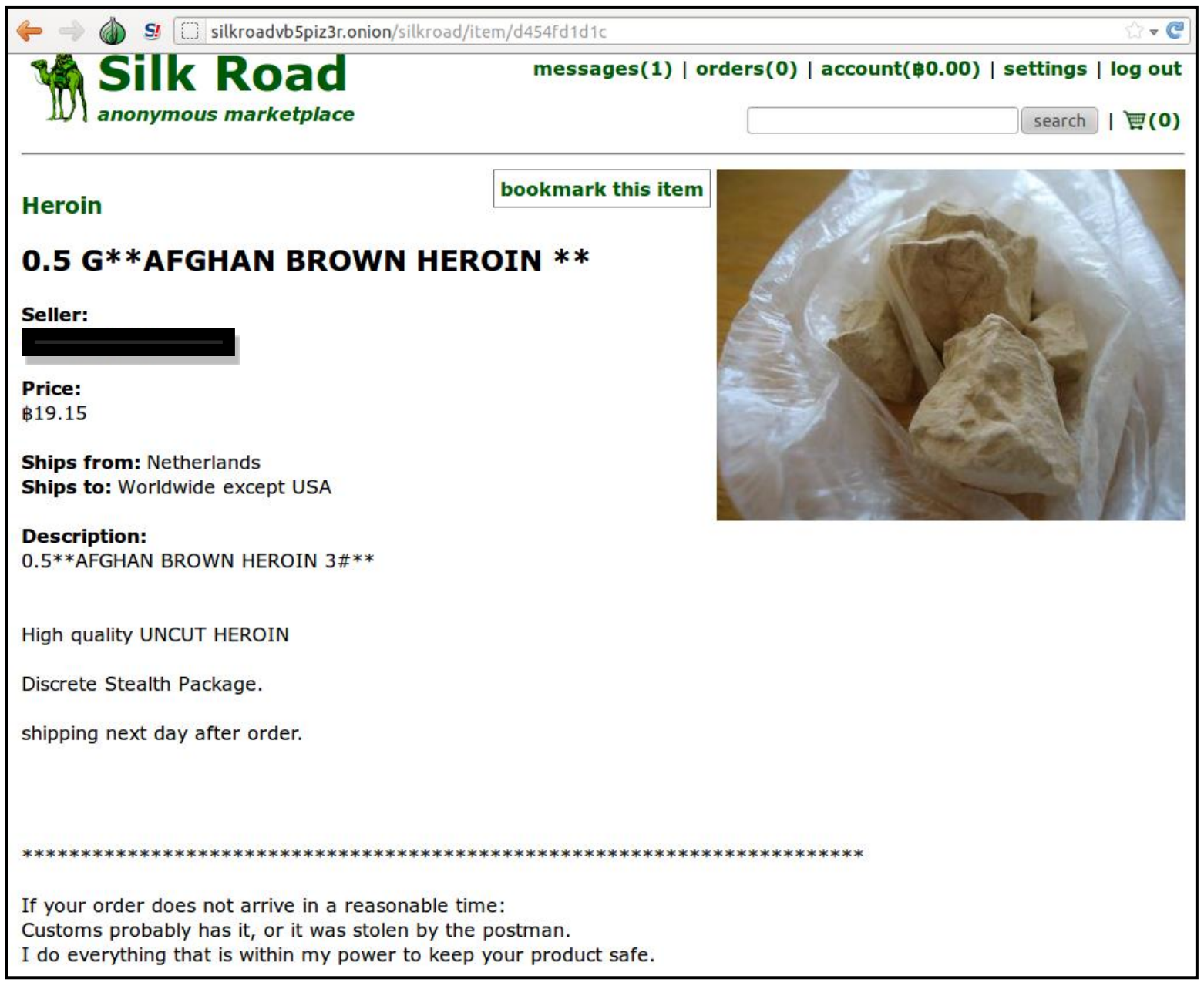

The BMPE is similar to *hawalas* – meaning transfer in Arabic – or *fe chi'en* in China: they are international underground banking systems. Online money laundering techniques are interconnected with these traditional methods, which are no longer 'localized'. Online operations contribute to complexity the audit process and thus hide the illegal source.

# Increasing trends in online money laundering

## Online Games as obfuscators

Online role playing games provide an easy way for criminals to launder money. This frequently involves the opening of numerous different accounts on various online games to move money.

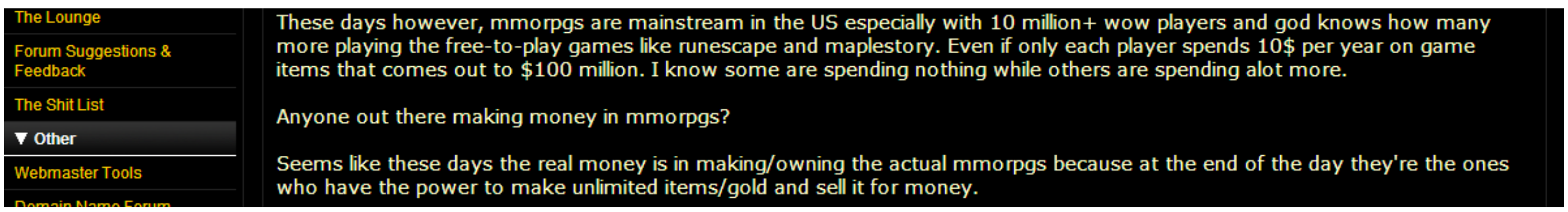

*Online games, an increasing source of interest in hacking forum boards*

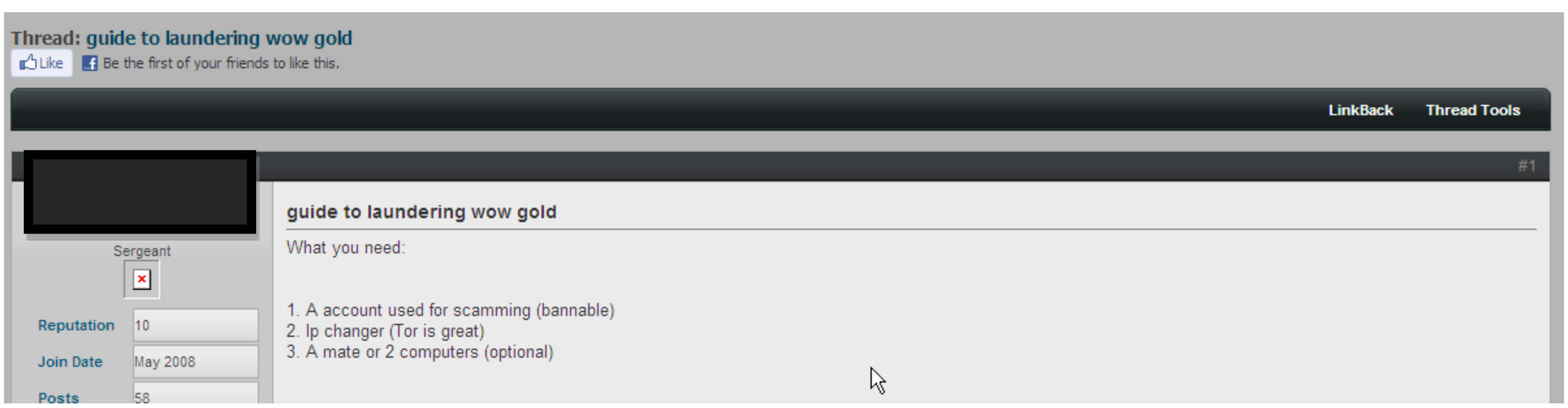

Other clever scammers have found a way to launder money using online games. Massively multi-player online role playing games (MMORPGs) provide an easy way for criminals to launder money. This frequently involves the opening of numerous different accounts on various online games.

Since MMORPGs use credits that players can exchange for real money, it is easy to do. Popular games for this type of scam include Second Life and World of Warcraft for instance.

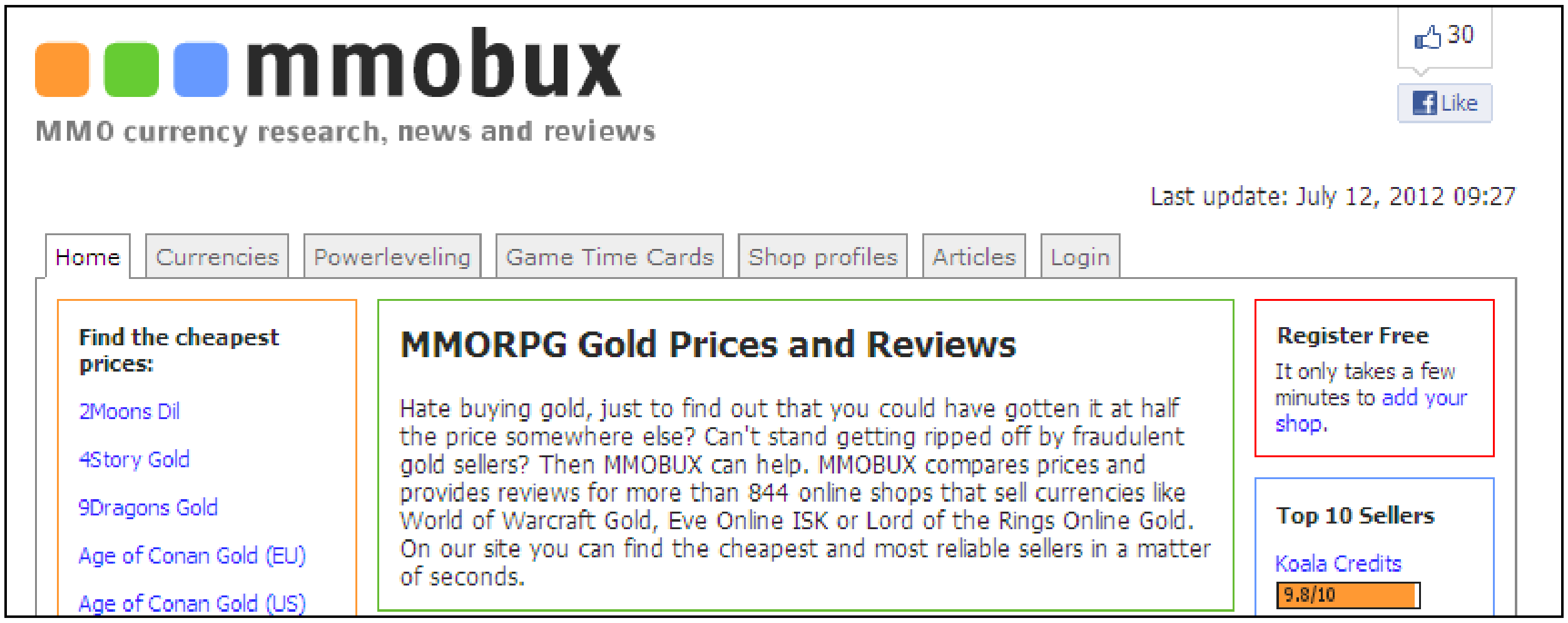

*While the market increase, new services emerge for gold sellers; those websites also provide prepaid cards useful for money laundering purposes*

MMORPG players very rarely know the people that they meet online. Citizens from dozens of different countries play these online games. Using the virtual currency systems in these games criminals in one country can send virtual money to associates in another country. Then, the virtual money can be transferred into real money, with the criminals leaving no trace of evidence authorities could follow back to them.

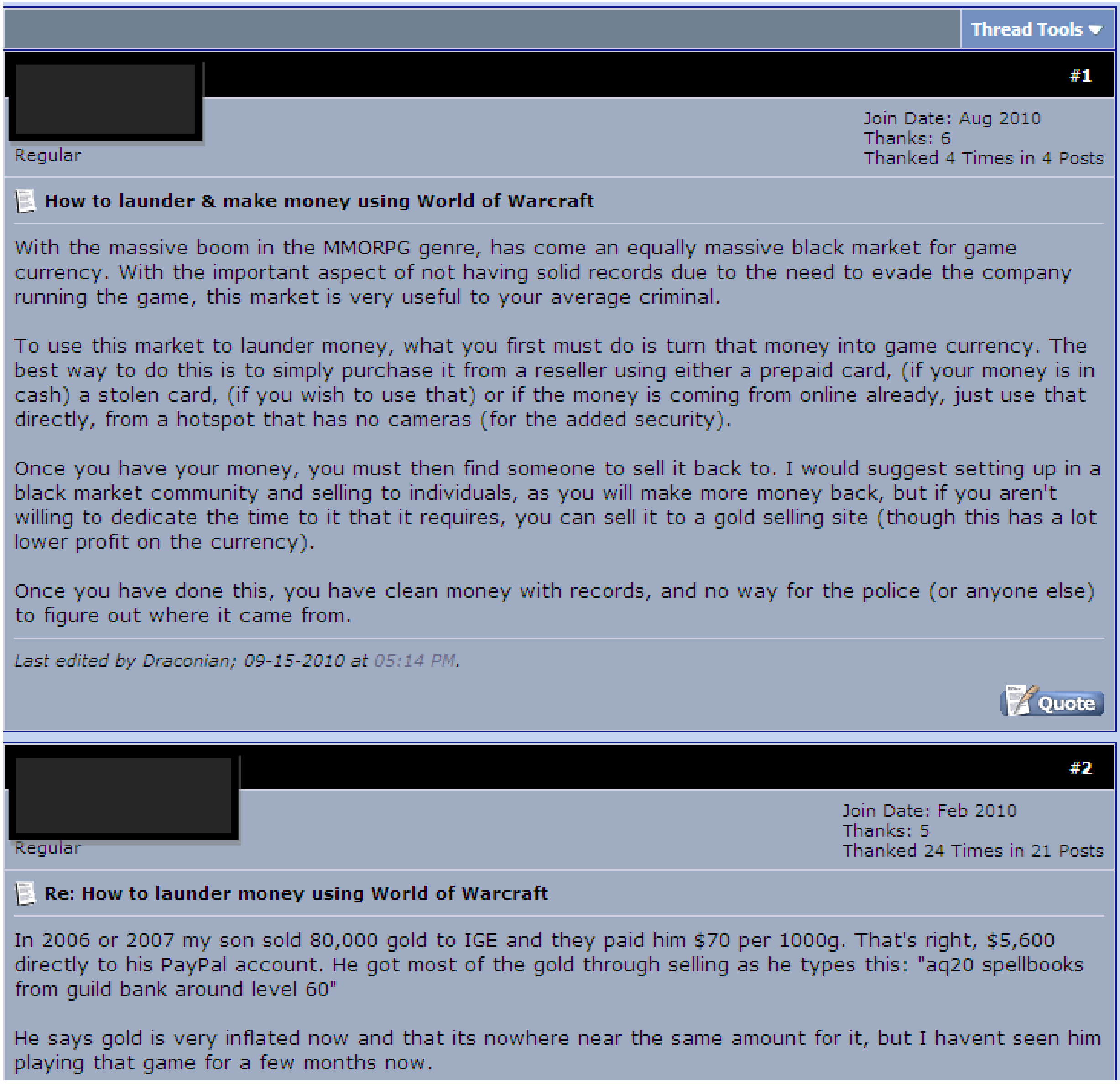

Thread Tools

[redacted] #1

Regular

Join Date: Aug 2010
Thanks: 6
Thanked 4 Times in 4 Posts

**How to launder & make money using World of Warcraft**

With the massive boom in the MMORPG genre, has come an equally massive black market for game currency. With the important aspect of not having solid records due to the need to evade the company running the game, this market is very useful to your average criminal.

To use this market to launder money, what you first must do is turn that money into game currency. The best way to do this is to simply purchase it from a reseller using either a prepaid card, (if your money is in cash) a stolen card, (if you wish to use that) or if the money is coming from online already, just use that directly, from a hotspot that has no cameras (for the added security).

Once you have your money, you must then find someone to sell it back to. I would suggest setting up in a black market community and selling to individuals, as you will make more money back, but if you aren't willing to dedicate the time to it that it requires, you can sell it to a gold selling site (though this has a lot lower profit on the currency).

Once you have done this, you have clean money with records, and no way for the police (or anyone else) to figure out where it came from.

*Last edited by Draconian; 09-15-2010 at 05:14 PM.*

Quote

[redacted] #2

Regular

Join Date: Feb 2010
Thanks: 5
Thanked 24 Times in 21 Posts

**Re: How to launder money using World of Warcraft**

In 2006 or 2007 my son sold 80,000 gold to IGE and they paid him $70 per 1000g. That's right, $5,600 directly to his PayPal account. He got most of the gold through selling as he types this: "aq20 spellbooks from guild bank around level 60"

He says gold is very inflated now and that its nowhere near the same amount for it, but I havent seen him playing that game for a few months now.

*Underground forums share money laundering best practices and tips*

It is important to note that the game developers are not associated with the scam at all, and in many cases are involved with the efforts to stop such crimes from occurring.

Recent years have seen the emergence of virtual communities and online gaming; scams related to MMORPGs will rise with these virtual worlds.

## Micro laundering: micro-payment, micro-jobs and m-commerce

Cyber criminals are increasingly looking at micro laundering via sites like PayPal or, interestingly, using job advertising sites, to avoid detection. Moreover, as online and mobile micro-payment are interconnected with traditional payment services, funds can now be moved to or from a variety of payment methods, increasing the difficulty to apprehend money launderers. Micro laundering makes it possible to launder a large amount of money in small amounts through thousands of electronic transactions. One growing scenario: using virtual credit cards as an alternative to prepaid mobile cards; they could be funded with a scammed bank account – with instant transaction – and used as a foundation of a PayPal account that would be laundered through a micro-laundering scheme.

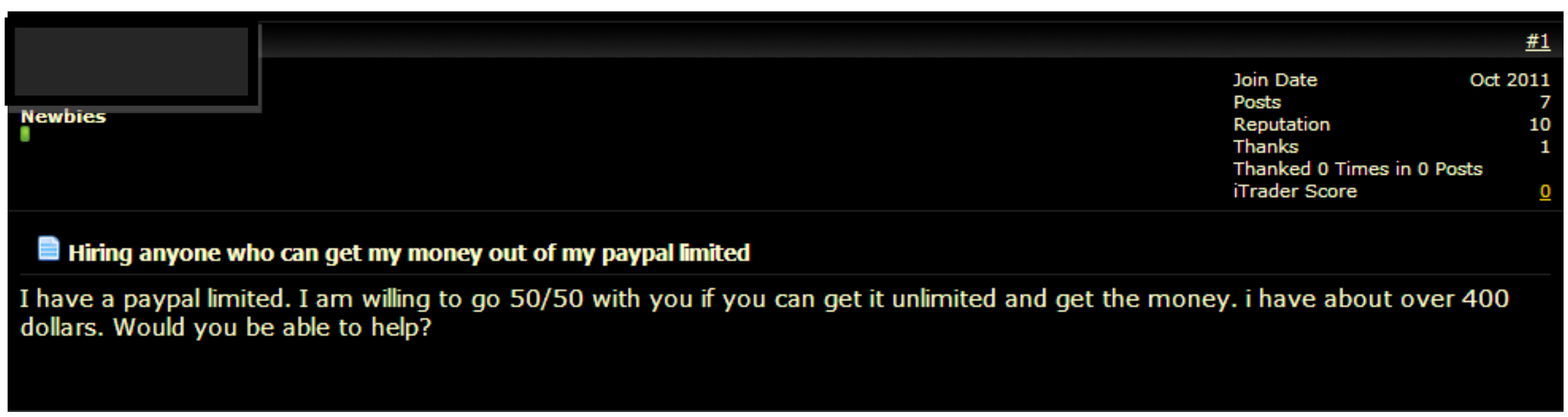

*Cleaning money might be an issue for some beginner money launderers… They frequently ask for partnership on forum boards.*

1. ***WHILE THE PAYPAL IS NOT YET LIMITED STILL IN GOOD STANDING***
*I send a friend a bunch of different invoices for amounts around $100 during the span of a week or 2.*

*****Reason Being when PayPal Limits you 80% of the time **you are still able to REFUND the $ thats in your account**. You refund to the friend that you had a transaction with in the past(60 Days), make him look like the victim, in the return message write, "I apologive for this inconvience, here is your money back"*

*. He then Withdrawls it, and sends you the money in whatever way, cash, Western Union, Money Gram, ETC...*

***. I call it *Ace in the Hole****

*#2*
*HOW TO GET MONEY OUT WITHOUT WITHDRAWING TO A BANK*

*Now this only works if your in good with a supplier, or someone overseas with a different jurisdiction, THE RECEIVER RISKS POSSIBLE LIMITATION TOO, WOULD USE STEALTH ACCOUNTS ON BOTH SIDES.*

*For me I used my Chinese supplier. Say I owed them $1500 for the number of shoes I had to order for the Day, and I had $2000 total ($500 Profit)*

*I would send the **whole 2K** and have them send me the $ Via **Western Union, or MoneyGram.** Paper trails are something u dont want. And this will **probably get fixed within days of me posting this.***

*I just know people are in tough spots, and All I want to do is help.*

*Thank me if you choose to.*

****** If your paypal is limited NOW, try refunding to someone **who u can trust to send the $ back to you**. It can be for a previous purchase.*

*For larger amounts of money, hackers' forum boards provide advice* on how to launder PayPal money while minimizing the commission

[redacted]

#12

Join Date: May 2011
Location: Where ever the wind blows
Posts: 673
Reputation: 56
Thanks: 69
Thanked 106 Times in 94 Posts

**Re: Best way to withdraw from PayPal?**

Why don't you open vendor account at say elance and pay yourself for the services. Of course this would mean opening or transfer fund to another paypal (perhaps your legit one) and then withdraw?

Either ways no matter what you do, with your set up, you are bound to lose a bit of commissions here and there.

**Reply With Quote**

**The Following User Says Thank You to seoguy81 For This Useful Post:**

[redacted] (05-24-2011)

*Laundering online money is a growing topic of interest on hackers' forum boards; in term of trends, the use of CCNOW is a favored PayPal alternative.*

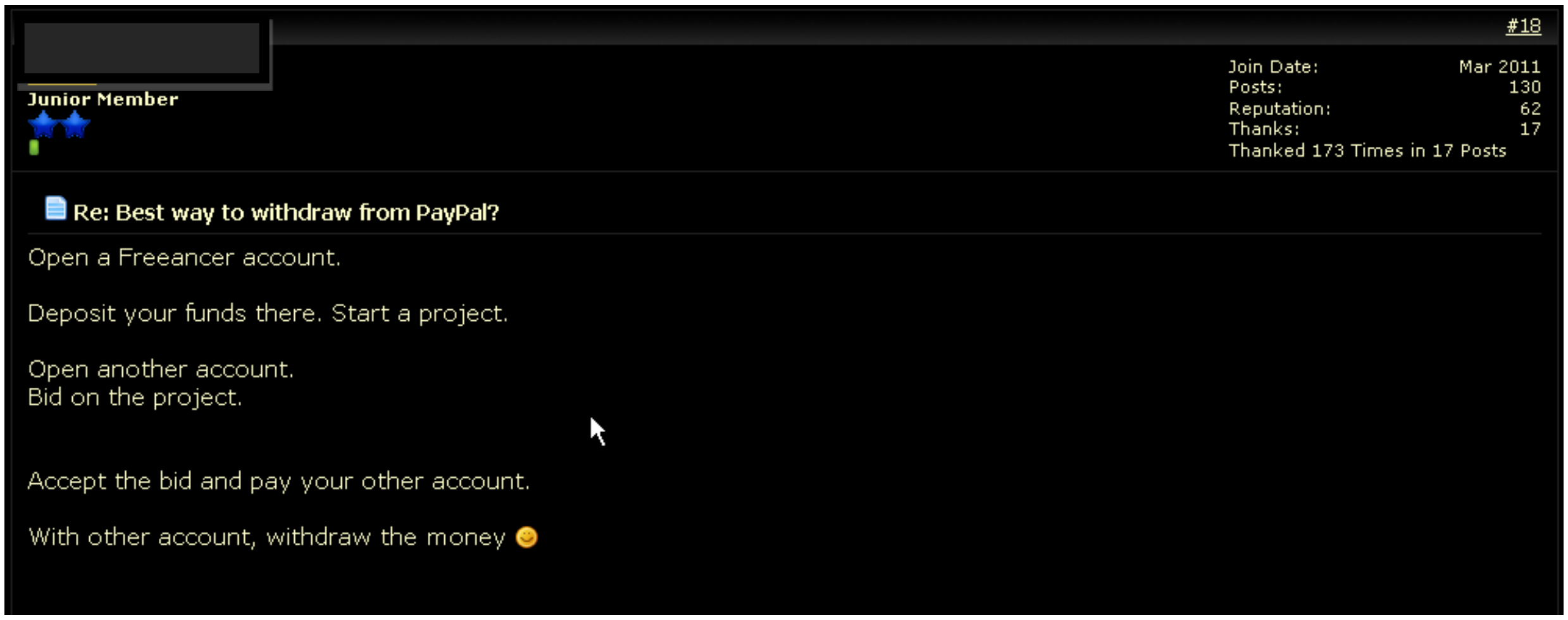

[redacted] #18

Junior Member

Join Date: Mar 2011
Posts: 130
Reputation: 62
Thanks: 17
Thanked 173 Times in 17 Posts

Re: Best way to withdraw from PayPal?

Open a Freeancer account.

Deposit your funds there. Start a project.

Open another account.
Bid on the project.

Accept the bid and pay your other account.

With other account, withdraw the money

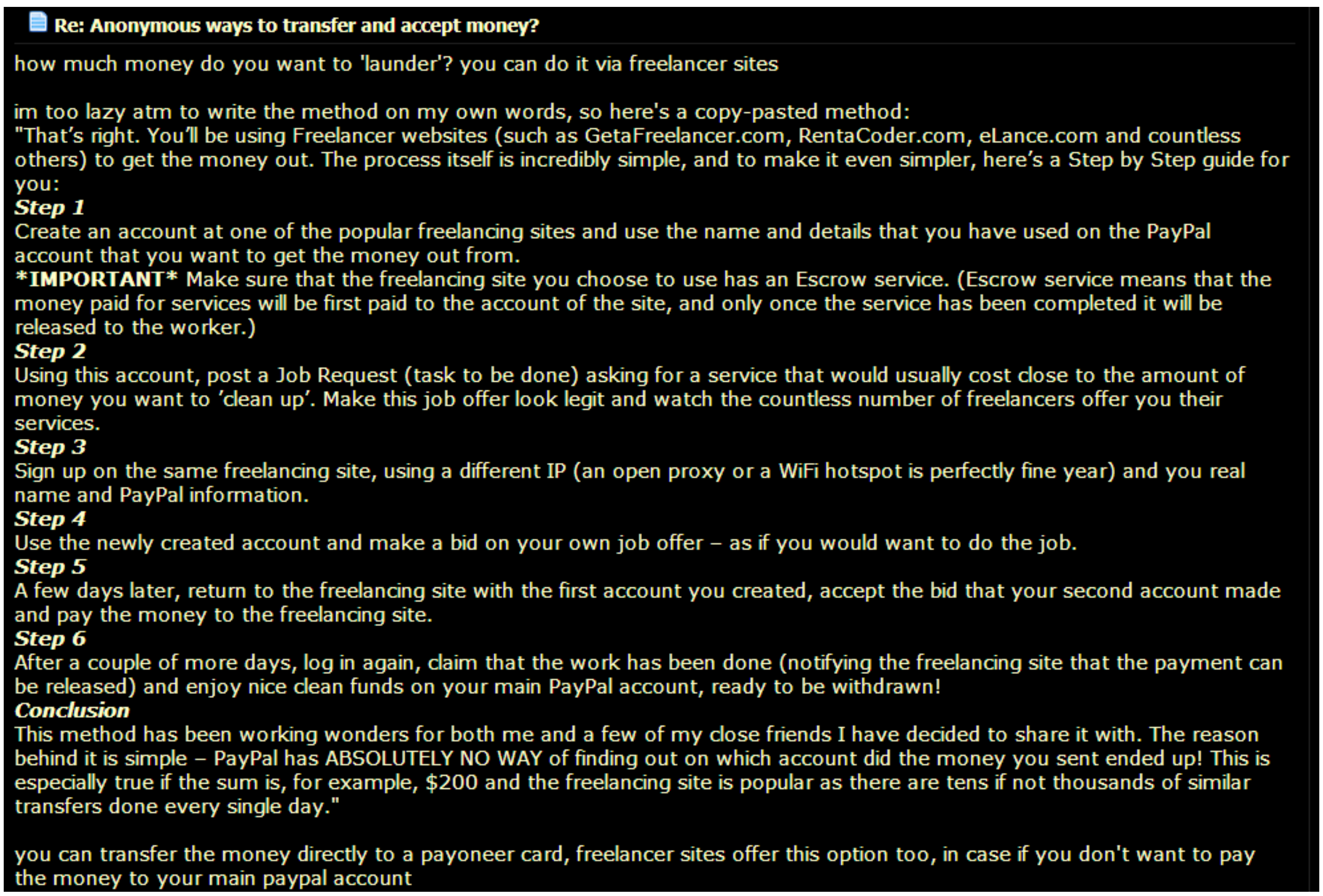

**Re: Anonymous ways to transfer and accept money?**

how much money do you want to 'launder'? you can do it via freelancer sites

im too lazy atm to write the method on my own words, so here's a copy-pasted method:
"That's right. You'll be using Freelancer websites (such as GetaFreelancer.com, RentaCoder.com, eLance.com and countless others) to get the money out. The process itself is incredibly simple, and to make it even simpler, here's a Step by Step guide for you:

***Step 1***
Create an account at one of the popular freelancing sites and use the name and details that you have used on the PayPal account that you want to get the money out from.
***IMPORTANT*** Make sure that the freelancing site you choose to use has an Escrow service. (Escrow service means that the money paid for services will be first paid to the account of the site, and only once the service has been completed it will be released to the worker.)

***Step 2***
Using this account, post a Job Request (task to be done) asking for a service that would usually cost close to the amount of money you want to 'clean up'. Make this job offer look legit and watch the countless number of freelancers offer you their services.

***Step 3***
Sign up on the same freelancing site, using a different IP (an open proxy or a WiFi hotspot is perfectly fine year) and you real name and PayPal information.

***Step 4***
Use the newly created account and make a bid on your own job offer – as if you would want to do the job.

***Step 5***
A few days later, return to the freelancing site with the first account you created, accept the bid that your second account made and pay the money to the freelancing site.

***Step 6***
After a couple of more days, log in again, claim that the work has been done (notifying the freelancing site that the payment can be released) and enjoy nice clean funds on your main PayPal account, ready to be withdrawn!

***Conclusion***
This method has been working wonders for both me and a few of my close friends I have decided to share it with. The reason behind it is simple – PayPal has ABSOLUTELY NO WAY of finding out on which account did the money you sent ended up! This is especially true if the sum is, for example, $200 and the freelancing site is popular as there are tens if not thousands of similar transfers done every single day."

you can transfer the money directly to a payoneer card, freelancer sites offer this option too, in case if you don't want to pay the money to your main paypal account

*Using online job marketplaces are an increasing recommendation*

Online jobs marketplaces such as Freelancer or Fiverr provide an interesting legitimate covert for money launderers. Job marketplaces could be used as anonymizers, escrow services providing a useful way for covering up the launderer's traces.

Fiverr micro-services ranging from $5 to $30 could also be used to reduce the footprint of a scammed account: amounts are small enough not to trigger a review under banking regulations. New money mule scams related to these platforms emerge. Indeed, many marketplaces users have been receiving emails asking them to join into a partnership involving the use of their accounts to sell services. They will usually provide some reason that they can’t open their own account, and offer to pay you a percentage of every transaction made through yours. These supposed contractors will put up many fake auctions, and ask you to transfer the funds from them to alias accounts. This is an evolution from the precedent model (classic money mule scam) as it no longer involves bank transfers but e-currency transfer (from Paypal to Bitcoin), and no longer concerns a large sum of money send once but small amounts send multiple times.

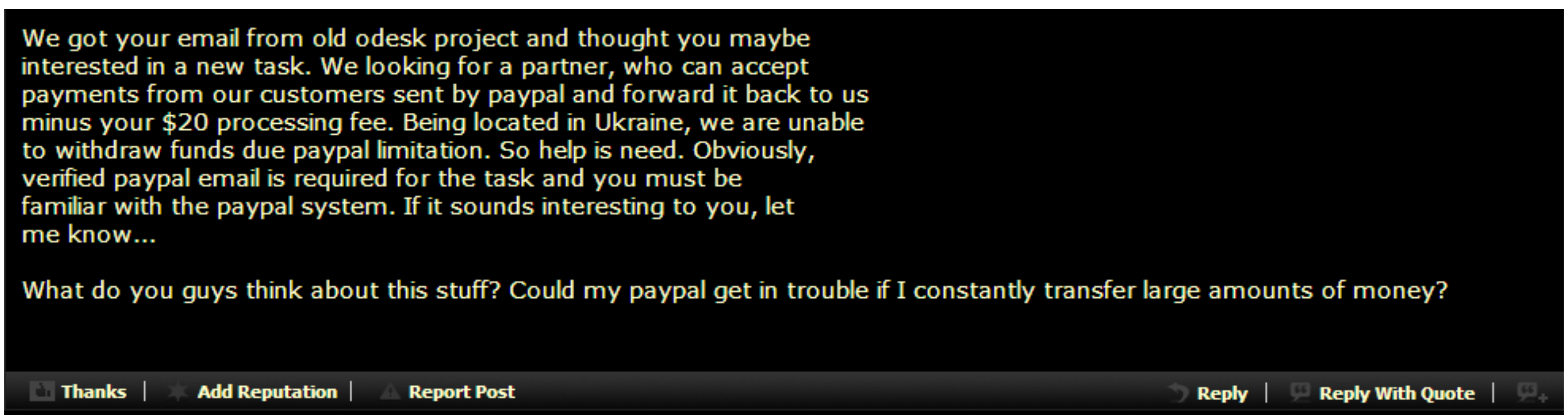
We got your email from old odesk project and thought you maybe interested in a new task. We looking for a partner, who can accept payments from our customers sent by paypal and forward it back to us minus your $20 processing fee. Being located in Ukraine, we are unable to withdraw funds due paypal limitation. So help is need. Obviously, verified paypal email is required for the task and you must be familiar with the paypal system. If it sounds interesting to you, let me know...

What do you guys think about this stuff? Could my paypal get in trouble if I constantly transfer large amounts of money?

Thanks | Add Reputation | Report Post    Reply | Reply With Quote |

The spread of mobile banking has been especially rapid and broad in Africa and Russia (see M-Pesa - Kenya's mobile wallet - and SMS Coin for instance) while global M-commerce market has grown steadily. It is now possible to send money from prepaid mobile card to criminal partners that will convert the credit into cash. This method provides criminals anonymity with a quick and easy access to money from nearly anywhere in the world.

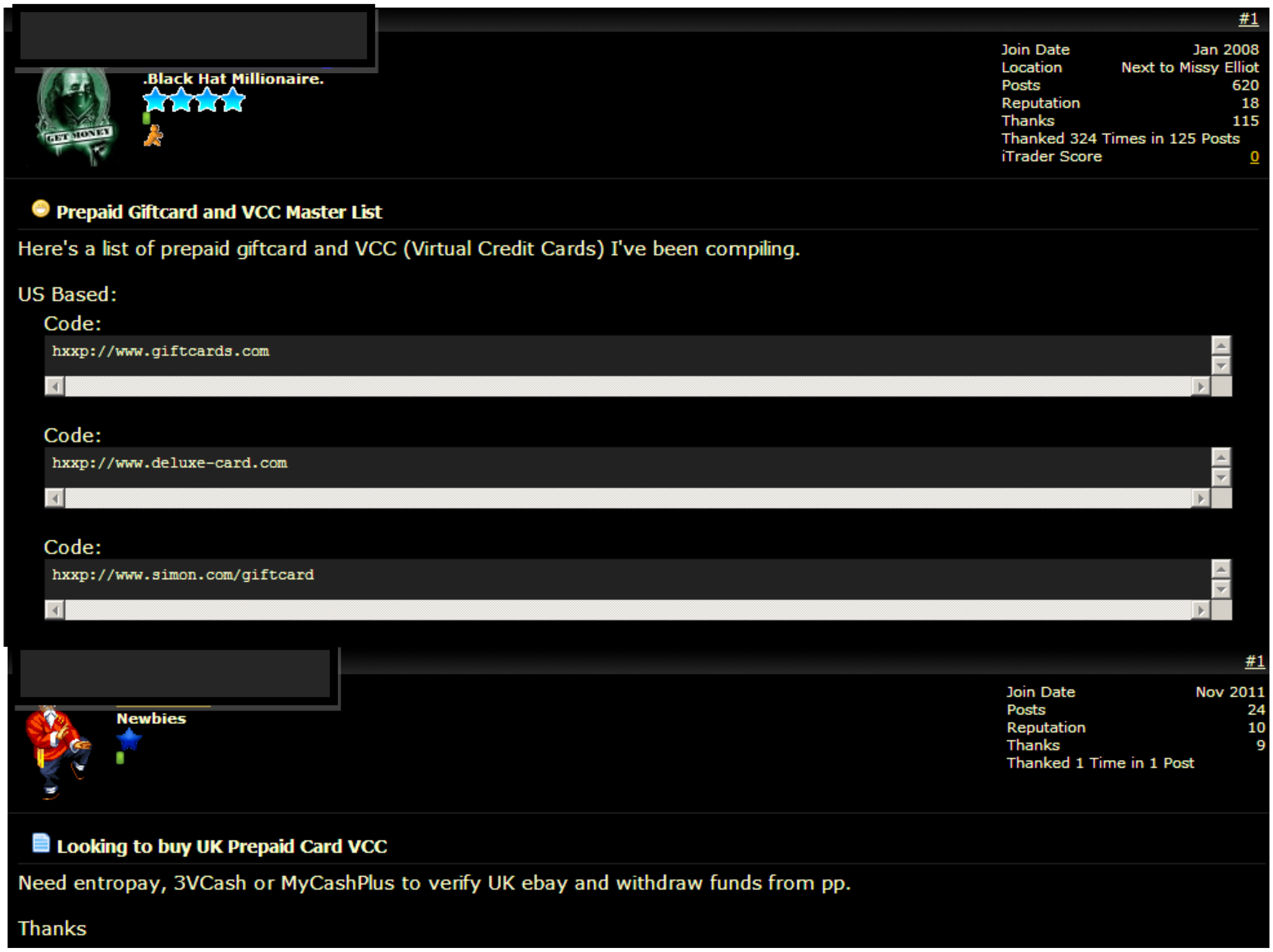

Virtual credit cards are a growing alternative to prepaid mobile cards; they could be funded with a scammed bank account – with instant transaction – and used as a foundation of a paypal account that would be laundered through a micro-laundering scheme.

Because online and mobile micro-payment are interconnected with traditional payment services, Funds can now be moved to or from a variety of payment methods, increasing the difficulty to apprehend money launderers. It is possible to launder a large amount of money in small amounts through thousands of electronic transactions.

# Conclusion

As we spend more time and money online, opportunities for criminals to involve us in their money laundering scams will only continue to grow. This will create an increasingly difficult situation for the various law enforcement agencies that are already being put to the test by the cunning of such criminals and the myriad untraceable means they have discovered to launder illegally obtained money.

As individuals, it is our responsibility to stay informed, and always be aware of the methods these criminals may use to involve us in their laundering schemes.

**Jean-Loup Richet**

## About the Author

Jean-Loup Richet is Information Systems Service Manager at Orange and Research Associate at ESSEC Business School - Institute for Strategic Innovation & Services. He graduated from the French National Institute of Telecommunications, Telecom Business School, and holds a research master's from IAE/HEC Paris.
Expert in IS Security, Jean-Loup Richet has been a speaker at several national and international conferences in Information Systems and has published articles in academic and trade journals. He is also a lecturer in IS Risk Management at Sorbonne Graduate Business School (International MBA).